\documentclass{pasa}%
\usepackage{graphicx}
\usepackage[dvipsnames, usenames]{color}

\usepackage{hyperref}

\usepackage{rotating}
\usepackage{gensymb}

\usepackage[utf8]{inputenc}
\usepackage[english]{babel}

\title[Thermonuclear burst reference catalogue]{Thermonuclear
burst observations for model comparisons: a reference sample}
\author[Galloway, Goodwin \& Keek]{Duncan K. Galloway$^{1,2}$, 
  Adelle J. Goodwin$^1$ \& Laurens Keek$^{3,4}$ \\
\affil{$^1$School of Physics \& Astronomy, Monash University, Clayton VIC
3800, Australia}%
\affil{$^2$also Monash Centre for Astrophysics (MoCA)}%
\affil{$^3$X-ray Astrophysics Laboratory, Astrophysics Science Division, NASA/GSFC, Greenbelt, MD 2077, USA}%
\affil{$^4$CRESST and the Department of Astronomy, University of Maryland, College Park, MD 20742, USA}}
\jid{PASA}
\doi{10.1017/pas.\the\year.xxx}
\jyear{\the\year}

\usepackage[authoryear]{natbib}
\bibpunct{(}{)}{;}{a}{}{,}
\newcommand{\gs}{GS~1826$-$24}
\newcommand{\sax}{SAX~J1808.4$-$3658}
\newcommand{\he}{4U~1820$-$303}
\newcommand{\su}{4U~1636$-$536}

\newcommand{\xte}{{\it RXTE}}
\newcommand{\epc}{{\rm erg\,cm^{-2}}}
\newcommand{\epcs}{{\rm erg\,cm^{-2}\,s^{-1}}}

\newcommand{\zcno}{Z_{\rm cno}}
\newcommand{\nat}{Nature}
\newcommand{\apj}{ApJ}
\newcommand{\apjl}{ApJL}
\newcommand{\apjs}{ApJS}
\newcommand{\aap}{A\&A}
\newcommand{\ssr}{SSR}
\newcommand{\mnras}{MNRAS}
\newcommand{\procspie}{Proc. SPIE}

\newcommand{\iaucirc}{IAUC}
\newcommand{\aaps}{A\&AS}

\newcommand{\saxdist}{$3.4\pm0.1$}
\newcommand{\saxx}{$0.48^{+0.12}_{-0.08}$}
\newcommand{\saxz}{$0.017^{+0.007}_{-0.005}$}
\newcommand{\saxdt}{$16.55\pm0.06$}

\newcommand{\gsdist}{6.1}
\newcommand{\gsopz}{1.23}
\newcommand{\gsrad}{12.1}   

\newcommand{\nhebursts}{67} 
\newcommand{\nheburstsxte}{16}

\begin{document}%
\begin{abstract}
We present a  sample of observations of thermonuclear
(type-I) X-ray bursts, selected 
for comparison with numerical models.
Provided are examples of four distinct cases of thermonuclear ignition:
He-ignition in mixed H/He fuel \cite[case 1 of][]{fhm81}; 
He-ignition in pure He fuel, following exhaustion of accreted H by steady
burning (case 2);
ignition in (almost) pure He accumulated from an evolved donor in an ultracompact
system; 
and 
an example of a superburst, thought to arise from ignition of a layer of carbon fuel produced as a by-product of more frequent bursts.
For regular bursts, we  measured the recurrence time and calculated averaged burst profiles from \xte\/ observations. We have also estimated the recurrence time for pairs of bursts, including those observed during a transient outburst modelled using a numerical ignition code.
For each pair of bursts we list the burst properties including recurrence time, fluence and peak flux, the persistent flux level (and inferred accretion rate) as well as the ratio of persistent flux to fluence. In the accompanying material we provide a bolometric lightcurve for each burst, determined from time-resolved spectral analysis. Along with the inferred or adopted parameters for each burst system, including distance, surface gravity, and redshift, these data are suggested as a suitable test cases for ignition models. 
\end{abstract}
\begin{keywords}
stars: neutron -- X-rays: bursts 
  -- astronomical databases: miscellaneous -- methods: numerical -- nuclear reactions 

\end{keywords}
\maketitle%
\section{INTRODUCTION }
\label{sec:intro}

Thermonuclear (type-I) bursts arise from unstable ignition of accreted fuel (typically mixed hydrogen and helium) on the surface of accreting neutron stars in low-mass binary systems.
Such events are of high priority for observers, as they provide information about the fuel composition, accretion rate, and even neutron star spin, mass and radius \cite[e.g.][]{heger07b,sb03,slb13}.
A key component contributing to our understanding of the burning physics is numerical modelling of the complex series of nuclear reactions which trigger and power the bursts \cite[e.g.][]{fhm81}.
To date, the degree to which predictions of numerical models have been compared in detail to observations is limited; this paper is part of a wider effort to address this situation.

The  physical conditions and processes which broadly influence the burst behaviour are relatively well understood.
The primary determinant is the local accretion rate $\dot{m}$, frequently expressed as a fraction of the 
Eddington rate $\dot{m}_{\rm Edd}$\footnote{Throughout this paper, we adopt a value of $\dot{m}_{\rm Edd}=8.8\times10^4\ {\rm g\,cm^{-2}\,s^{-1}}$}, at which the outwards 
{force due to} 
radiation pressure (assuming spherical symmetry) equals the surface gravity. 
The Eddington rate is also the predicted threshold at which the burning is expected to stabilise \cite[e.g.][]{ramesh03}, although observationally bursting behaviour appears to cease for many sources at a substantially lower level \cite[]{corn03a,bcatalog}.

For systems accreting a mix of hydrogen and helium
{at typical accretion rates ($\sim0.1\dot{m}_{\rm Edd}$)}, the fuel layer is  hot enough that H will burn stably via the hot-CNO cycle. This process is limited by $\beta$-decay, and will exhaust the H at the base of the layer in a time
\begin{equation}
    t_{\rm CNO} = 9.8 \left(\frac{X_0}{0.7}\right)\left(\frac{\zcno}{0.02}\right)^{-1}\ {\rm hr}
\end{equation}
\cite[e.g.][]{lampe16} which depends only on the accreted H-fraction $X_0$ and the metallicity $\zcno$.
Thus,
the composition of the fuel layer at ignition may 
vary, and a number of categories have
been identified by numerical studies. If the recurrence time is short, the burst will ignite in a mixed H/He environment, leading to a 
``case 1'' \cite[]{fhm81} or ``prompt mixed'' \cite[]{ramesh03} burst.
At lower accretion rates, the recurrence time may exceed $t_{\rm CNO}$, leading to ignition in a pure-He layer, i.e. ``case 2'' or ``delayed helium'' bursts.
At even lower rates, it is thought that H-burning too will become unstable, leading to ignition via that mechanism; but no unambiguous examples of such bursts have been observed.

An additional ignition case is provided by neutron stars in ultracompact systems, which likely accrete (almost?) pure He. Here the ignition will always be in a 
{H-poor}
environment, largely independently of the accretion rate.

The nuclear energy generation during the burst is related to the average
H-fraction in the fuel $\left<X\right>$, with $Q_{\rm
nuc}=1.6+4\left<X\right>\ {\rm MeV\,nucleon^{-1}}$.
The composition of the fuel may be inferred from the $\alpha$
parameter, the ratio of burst energy to accretion energy over the burst
recurrence time. For a neutron star of mass $M$, radius $R$, 
\begin{equation}
    \alpha = \frac{Q_{\rm grav}}{Q_{\rm nuc}}(1+z) \approx \frac{GM}{R} Q_{\rm nuc}^{-1} (1+z)
\end{equation}
{where $z$ is the gravitational redshift at the neutron star surface, given by $1+z=(1-2GM/Rc^2)^{-1/2}$. }
Provided the accretion rate is proportional to the persistent flux $F_p$, 
$\alpha$ may be measured (up to a factor corresponding to the ratio of the anisotropy of burst and persistent emission) as
\begin{equation}
    \alpha_{\rm obs} = \frac{F_p c_{\rm bol} \Delta t}{E_b} 
\end{equation}
where $c_{\rm bol}$ is the bolometric correction to the (band-limited) flux $F_p$, $\Delta t$ is the (regular) burst recurrence time, and $E_b$ the burst {bolometric} fluence (total energy).

The composition of the fuel may also be deduced from the shape of the lightcurve. He burns during the burst via the triple-$\alpha$ reaction, which proceeds on a much faster timescale than the hot-CNO, rp- and $(\alpha,p)$ reactions which burn H. The higher the He mass fraction, the faster the burning will proceed during the burst, reflecting in the burst rise and duration, as well as the timescale $\tau$ (the ratio of the fluence to the peak flux).

The numerical codes that have been used to compare to observations fall into three broad classes. The first class determines the ignition conditions given the accretion rate and fuel composition, and predicts the burst recurrence time and energetics. While this model has been developed to compare with the 
inferred atmospheric expansion 
during a burst as suggested by measurements of burst oscillations \cite[]{cb00}, it has also been used to compare with observations of bursts at low accretion rates \cite[]{gal06c}.
The primary disadvantage of such models is that they do not follow the time-dependent compositional structure of the atmosphere, which is significantly modified by the thermonuclear burning. 

The next class of models are one-zone time-dependent codes that simulate some fraction of the nuclear reaction network. Such codes have been used to demonstrate the extent of the rp-process that powers mixed H/He bursts \cite[]{schatz01}, as well as serving to probe the sensitivity of burst lightcurves to individual reaction rates \cite[]{cyburt16}.

The current (practical) state of the art for burst modelling is in 1-D multi-zone models that track the full extent of the nuclear reaction networks. Codes in this class include {\sc kepler} \cite[]{woos04}; the general relativistic hydrodynamics code {\sc AGILE},  coupled with a nuclear reaction network \cite[]{fisker06}, and {\sc MESA} \cite[]{mesa15}.

{1-D multi-zone models } have been used to make the most detailed comparisons of observations to models, most notably for \gs\ \cite[]{heger07b,zamfir12a}.
However, these comparisons are generally limited in the extent of the sources and the observational data utilised. These limitations, despite the extensive modeling capabilities (and accumulated observational data) motivate the present study.

{
Here we present a sample of observations of thermonuclear bursts, intended for comparison with numerical  models. The objectives of the sample assembly are twofold: first, as test cases to quantify variations between different codes (and hence intrinsic model uncertainties); and second, as examples that may be used to refine the system parameters by direct comparison with individual codes.

We deliberately take a pragmatic approach which incorporates constraints on system parameters obtained by comparison with particular numerical models (of varying degrees of fidelity). While this approach introduces dependence of the system parameters on the specific models chosen for these analyses, this dependence is unimportant for either of the two objectives described above. 
For the code comparisons, the specific choice of input parameters does not matter, provided they are  plausible; and the parameters we have adopted are the best current parameter estimates for the systems studied.
As for improving the system parameters via more detailed comparisons with individual models, the adopted system parameters provided here should be viewed as a suggested starting point. 
}

This paper is presented as follows. In section \S\ref{sources} we describe the sources selected to make up the sample. 
In \S\ref{observations} we describe the observational data from which the sample lightcurves and other parameters are drawn.
In section \S\ref{results} we briefly describe the inferred properties of the bursting sources, and compare the lightcurves in the sample.
Finally, in \S\ref{discussion} we suggest how the assembled data may be applied and used to test numerical models.

\section{SOURCE SELECTION}
\label{sources}

We selected four well-studied burst sources, which span the range of
theoretical ignition cases identified observationally (Table \ref{tab:sources}).
We also selected a fourth source to serve as an example of a superburst, \su, which is notable 
{as one of only two such events}
observed at high sensitivity with \xte\/ \cite[e.g.][]{keek14a}.

{We  list in Table \ref{tab:sources} }
those quantities required to simulate thermonuclear bursts and compare to the observations, including the distance, mass fraction of hydrogen  $X_0$ and of CNO nuclei, $Z_{\rm CNO}$, and the adopted redshift $1+z$ and surface gravity $g$. 
We caution that in most cases these parameters are not precisely known, however for the purposes of deciding upon suitable values for simulation tests, we adopt values which are generally consistent with the observed burst properties and/or the neutron star population. 

The surface gravity and fuel composition parameters are key input to the models, while the redshift is used to transform the burst lightcurves (and recurrence times) from the neutron-star frame to the observer's frame (see \S\ref{discussion}).
The source distance $d$ 
{
is required to convert model predictions to observed quantities, and
is subject to specific uncertainties depending upon the method by which it is estimated, as described further in 
appendix \ref{distance}.
}

\begin{figure*}
\begin{center}
\includegraphics[width=17.5cm]{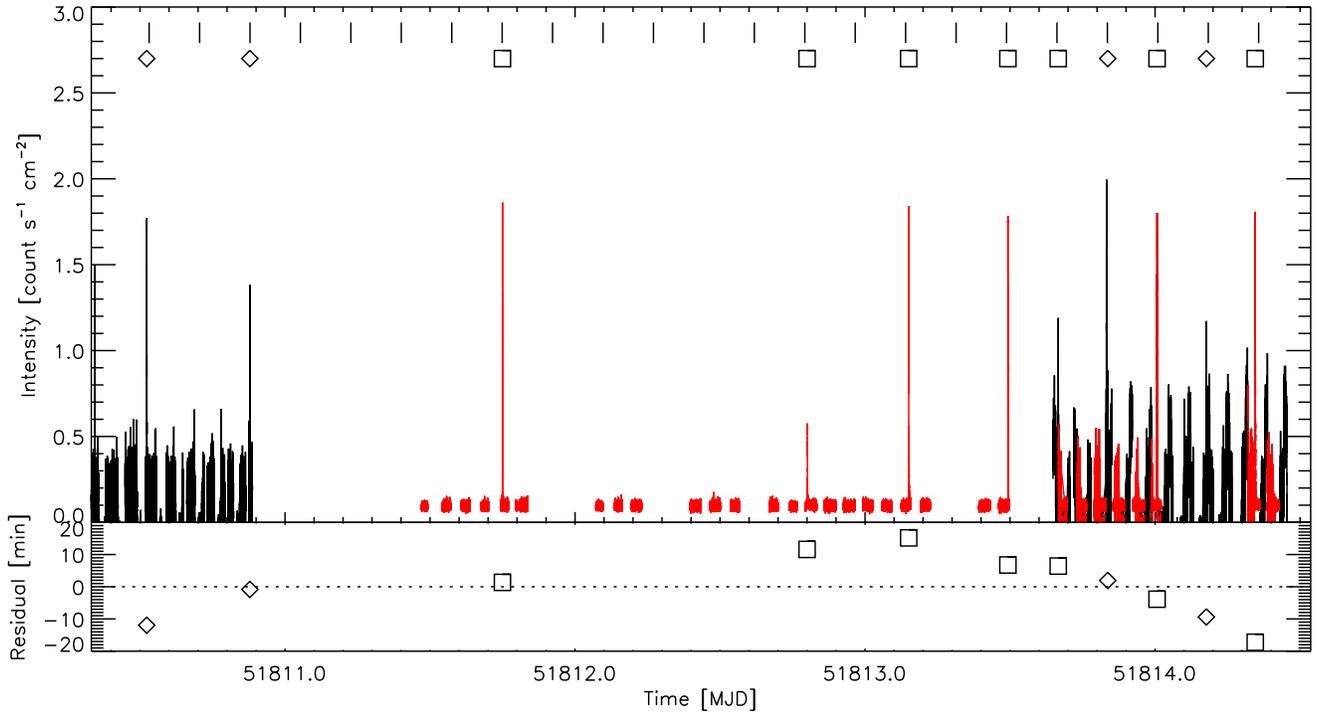}
\caption{Example lightcurve showing bursts observed in 2000 September by {\it BeppoSAX}/WFC and the {\it Rossi
X-ray Timing Explorer}/PCA from \gs.
The top panel shows the lightcurves from the two instruments, rebinned to 10-s resolution, with the detected bursts marked with  open symbols. The WFC lightcurve is shown in black, with open diamonds indicating the bursts; for PCA, the  lightcurve is red, with open squares marking the bursts. 
The vertical lines above the symbols indicate the predicted times of bursts according to a constant recurrence time model; note that the unobserved bursts in the model train consistently fall in the data gaps.
In the lower panel we show the residual to the constant recurrence time model. Here the best-fit recurrence time is $4.177\pm0.010$~hr, with an RMS error of 9.53~min.
}\label{example}
\end{center}
\end{figure*}

\begin{table*}[h]
\begin{center}
\caption{Target thermonuclear burst source properties}\label{tab:sources}
\begin{tabular}{lccccccl}
\hline    
   & Dist.  &  \multicolumn{2}{c}{Accreted fuel} & & $g$ & $R$ \\
Source  & (kpc)  & $X_0$ & $Z_{\rm CNO}$ & $1+z$ & ($10^{14}\,{\rm cm\,s^{-2}}$) 
& (km) & Ref. \\
\hline           
\gs &  \gsdist &  0.7 & 0.02 & \gsopz & {\it 2.34} & \gsrad & [1,2] \\
\sax  &  \saxdist & \saxx & \saxz & {\it 1.26} & {\it 1.86} & {\it 11.2} & [2]\\
\he &  $7.6\pm0.4$  & $\lesssim0.1$ & {\it 0.02} & 1.409 & 2.96 & $11.1\pm1.8$ & [3,4,5] \\
\su  & $5.6\pm0.4$  & {\it 0.7} & {\it 0.02} & {\it 1.26} & {\it 1.86} & {\it 11.2} \\
\hline
\end{tabular}
\medskip\\
Values in italics indicate there are no  constraints specific for that source. \\
References:
1. \cite{heger07b}; 2. this work; 
3. \cite{kuul03a}; 4. \cite{cumming03}; 5. \cite{ozel16}
\end{center}
\end{table*}

Below we describe the basic observational properties of each source.

\subsection{\gs}
\label{s1826}

This source, discovered as a new transient by {\it Ginga} in 1988
September \cite[]{tanaka89}, has been X-ray bright and exhibiting bursts
ever since.
It is known as the ``clocked'' or ``textbook'' burster, due to its
unusually regular and consistent bursts \cite[]{clock99,gal03d}. 
The inferred accretion rate varies little on short timescales, and is within the range 5--13\%~$\dot{m}_{\rm Edd}$ \cite[]{bcatalog,chenevez16}.
This pattern of regular bursting behaviour has been broken only recently, with the transition to weaker, irregular bursts accompanied by a soft spectral state, in 2014 June \cite[]{chenevez16}.
By virtue
of the typically consistent bursting behaviour, \gs\ is also one of the few sources for which detailed
efforts have been made to deduce the composition of the accreted fuel.
However, these efforts have not yet resulted in unambiguous results.

The measured $\alpha$-values for the source, at $\approx40$, strongly
suggest a high fraction of H in the burst fuel, and so we adopt the solar value, i.e. $X_0=0.7$, for the accreted H fraction.
The lack of variation in burst fluence (i.e. total burst energy) over a
range in accretion rate was taken as evidence by \cite{gal03d} that the
fuel must be deficient in CNO nuclei. Otherwise, steady (hot-CNO)
H-burning between the bursts would signficantly change the fuel
composition at ignition, as the burst recurrence time decreased from
around 6~hr to 4~hr.
A subsequent comparison of burst properties, including the detailed shape
of the lightcurve, with time-dependent 1-D model predictions by {\sc
kepler} \cite[]{woos04} led instead to the conclusion that the accreted
fuel must be approximately solar in composition, i.e. $Z_{\rm CNO}\approx0.02$, 
at odds with the earlier
analysis \cite[]{heger07b}. Since {\sc kepler} includes substantially more physics than the code used in the earlier analysis, we adopt solar composition following the most recent study.

Despite the uncertainty in the fuel composition, the properties of the
bursts from \gs\ strongly suggest burning of mixed H/He fuel by unstable
He ignition, i.e. ``case 1'' of \cite{fhm81} (or alternatively, ``prompt
mixed'' bursts in regime 3 of \citealt{ramesh03}).

\subsection{SAX J1808.4$-$3658}
\label{s1808}

This source was discovered as a new transient in 1996 with {\it BeppoSAX}
\cite[]{zand98c}, and has been active every 2--3 years ever since.
Observations made during the 1998 outburst by \xte\/ revealed
401~Hz X-ray pulsations, making it the first accretion-powered
millisecond pulsar \cite[]{wij98b}. Bright, radius-expansion bursts have
been observed in almost every outburst, and the detection of burst
oscillations also at 401~Hz in bursts during the outburst of 2002 October
confirmed the link between oscillation frequency and neutron-star spin
\cite[]{chak03a}.

The transient outbursts typically reach a maximum inferred accretion rate
of only $0.05\ \dot{m}_{\rm Edd}$, and the bursts 
have fast ($\lesssim0.5$~s) rises indicative of
He-rich fuel.
The wide (2.01~hr) orbit can accommodate a (H-rich)
companion \cite[]{chak98d}, and it is thought that the accreted H is burned away by $\beta$-limited hot CNO burning prior to ignition.
The $\approx20$--30~hr recurrence times, given the low accretion rates, are sufficient provided the  metallicity, $\zcno$ is not far below the solar value.
This type of burst corresponds to ``case 2'' of \cite{fhm81}, or ``delayed helium'' (regime 4) of 
\cite{ramesh03}.

Comparisons of the burst properties with the predictions of numerical
ignition models indicate a range of possible compositions, with
$\zcno\propto X$ \cite[]{gal06c}. The lowest plausible H-fraction in the
accreted fuel was $X_0\approx0.35$.

\subsection{3A 1820$-$303}
\label{s1820}

Located in the globular cluster NGC~6624, \he\ (also known as Sgr X-4) was first detected as a bursting source with the {\em Astronomical Netherlands Satellite}\/ \cite[ANS;][]{grindlay76}. When in the bursting state, the source exhibits quasi-regular radius-expansion bursts with recurrence times in the range 2--4~hr \cite[e.g.][]{cumming03}. The source intensity is modulated on a $\approx176$-d periodicity \cite[]{pt84}, and it only exhibits bursts in the (relatively short) low-intensity phase of the cycle, when the inferred accretion rate is $\approx0.2$--0.95~$\dot{M}_{\rm Edd}$ \cite[e.g.][]{bcatalog}.

The orbital period of 685~s is known from periodic modulation of both persistent X-rays and the optical counterpart \cite[]{kw86,swp87}, and indicates an ``ultracompact'' binary which is too close for a Roche-lobe filling main-sequence star. The mass donor is thus assumed to be a very low-mass He white dwarf, and thus accreting pure He (i.e. $X_0=0$).
Some evolutionary models predict a small ($X\sim 0.1$) amount of H in the surface layers, and this cannot be ruled out from comparing the observed bursts with ignition models \cite[]{cumming03}.

The mass and radius of the neutron star in this system have been estimated by equating the blackbody normalisation in the cooling tail and the peak flux of PRE bursts, and adopting the inferred cluster distance \cite[]{ozel16}.
Although unresolved systematic errors may yet affect such measurements \cite[e.g.][]{slb13}, these values represent the best current estimates for the source, and are adopted in Table \ref{tab:sources}.


\subsection{4U 1636$-$536}
\label{s1636}

\su\ is a persistently-accreting burst source in a 3.8~hr orbit \cite[]{1636orb}. The system is one of the most prolific type-I (thermonuclear) bursters known, and has been studied extensively with most major observatories \cite[e.g.][]{lew93,bcatalog}. The neutron star spin has been measured from burst oscillations and transient pulsations during a superburst, at 579.3~Hz \cite[]{zhang97,stroh02b}. The relatively wide orbit, spectral features from hydrogen \cite[]{august98}, and the (at times) long profiles of the bursts strongly suggest the mass donor is a main-sequence star which accretes hydrogen-rich fuel.

The system has exhibited long-term variations in its persistent intensity. Between 1996 and 2001, the X-ray flux was in the range 3--$6\times10^{-9}\ \epcs$, indicating an accretion rate of 11--21\%~$\dot{m}_{\rm Edd}$ \cite[cf. with][]{bcatalog}. Since then, the flux has been consistently approximately 50\% lower. At these accretion rates, it would be expected that bursts would consistently exhibit long profiles arising from ignition of mixed H/He fuel, although this is not always the case. \su\ is also well-known for short-recurrence time bursts occuring in trains of up to 4 \cite[]{keek10}.

The first superbursts were detected from \xte/ASM data \cite[]{wij01b}, and two additional events were subsequently reported, one also detected by the PCA \cite[]{kuul04,kuul09b}. All the superbursts were detected while the system was in it's higher flux range, prior to 2001. 
The 2001 event, on February 22 (MJD 51962.7),  was observed with the PCA instrument, and has offered the highest signal-to-noise data during such an event, comparable only to the event seen from \he\ \cite[]{stroh02}. Preliminary spectral fitting of the \su\ burst indicated that the spectrum could not be adequately fitted with blackbody models \cite[]{kuul04}, and subsequently it was shown that the burst spectrum included substantial contributions arising from reflection of the burst emission from the accretion disk \cite[]{keek14a}.

\section{OBSERVATIONS \& ANALYSIS}
\label{observations}

{
We used preliminary results from the  Multi-INstrument Burst ARchive
(MINBAR\footnote{{\url http://burst.sci.monash.edu/minbar}}), which includes
bursts observed by the {\it Rossi X-ray Timing Explorer} Proportional Counter Array (\xte/PCA), {\it BeppoSAX} Wide-Field Camera \cite[WFC;][]{jager97,zand04b} and {\it
INTEGRAL} Joint European X-Ray Monitor \cite[JEM-X;][]{lund03}.

We calculated burst lightcurves from time-resolved spectroscopic analysis of  selected bursts observed by \xte/PCA \cite[]{xte96}. The MINBAR analysis follows the approach of \cite{bcatalog}, incorporating the latest PCA responses and the effects of deadtime, as for the measured peak PRE burst fluxes (see appendix \ref{distance}).
Earlier analyses of samples including these data have demonstrated evidence for an increased contribution of persistent flux during bursts, possibly arising from the effects of Poynting-Robertson drag on the inner accretion disk
\cite[]{worpel13a,worpel15}. While the fractional increase in the persistent flux can be $\sim10$, this contribution is a small fraction of the burst flux, and so any correction required to the burst flux is also small (of order a few per cent, similar to the intrinsic uncertainty of the time-resolved flux measurements). Thus, we neglect this effect for the purposes of determining our mean profiles.
}

\subsection{Selecting burst sequences}

For \gs\ and \he, we
identified burst {sequences} in MINBAR for which we
could reliably infer the burst recurrence time, even when the observations
were interrupted. Such interruptions arise for instruments in low-Earth
orbit, due to Earth occultations and passages through the South Atlantic
Anomaly. An example burst train is shown in Figure \ref{example}.

We then selected and averaged bursts observed by the \xte/PCA  within each burst train. MINBAR
also includes bursts observed by {\it BeppoSAX}/WFC and {\it
INTEGRAL}/JEM-X, which are useful for establishing the regularity of the bursts,
but the sensitivity of these instruments is much lower
than \xte, and so are unsuitable for producing high-quality burst
lightcurves. 

For \gs, we  augmented our burst sample with optical bursts observed
during \xte\/ observations in 1998 June, as also analysed by
\cite{thompson08}. 

We measured the average recurrence time of each train by assigning a trial
integer value to each burst, and then performing a linear
least-squares fit to the times. The trial values were determined by
dividing the time since the first burst, by the minimum burst separation
of the train; we subsequently adjusted the sequence numbers to minimise
the RMS value of the fit. We typically obtained an RMS value of
$\approx20$~min or better. 

From the set of bursts in each train $N_{\rm burst}$, we then selected the
\xte\/ bursts ($N_{\rm av}$, { Table \ref{tab:bursts}}), and created average lightcurves from each
train.
We also calculated averaged burst fluences, peak fluxes and $\alpha=\Delta t F_p c_{\rm bol}/E_b$, where $F_p$ is the persistent (accretion) flux, $c_{\rm bol}$  the bolometric correction (see \S\ref{bolcorr}), and $E_b$ the burst fluence.

We estimated the accretion rate $\dot{m}$ as a fraction of the Eddington rate $\dot{m}_{\rm Edd}$, 
as
\begin{equation}
\dot{m}=4\pi d^2 F_pc_{\rm bol}/\dot{m}_{\rm Edd}
\end{equation}
where $F_p$ is the average persistent flux over the burst train, and $d$ the estimated source distance from Table \ref{tab:sources}.
This estimate neglects the possible anisotropy of the persistent emission, so must be taken as  a guide only.

The identification of suitable burst trains for \he\ was made  difficult by the relatively small number of bursts available in the MINBAR sample. Just \nhebursts\ bursts were detected by {\it BeppoSAX}/WFC, {\it INTEGRAL}/JEM-X or \xte/PCA, with only \nheburstsxte\ events observed with \xte.
We ultimately identified eight burst trains with three or more bursts and evidence for regular bursting behaviour; the range of inferred recurrence time was 2.7--4.4~hr. 

Only two of these sequences included bursts detected with \xte, each with just three bursts. For each sequence, we observed a pair of bursts within a few hours of each other, and a third burst some days earlier or later. We ultimately rejected one of these sequences, in 2009 May \cite[MJD~54980; see also ][]{zand12a} 
because there were significant variations in the persistent flux (and spectral shape) over the interval, and it was not possible to be confident about the recurrence time for the close pair of bursts, due to data gaps between them. 
This was not the case for the remaining train, in 1997 May (MJD~50572), and so we retained this set of bursts. With only one burst observed by \xte\/ in this sequence, we simply provide the time-resolved spectroscopy results from that burst, as was done for \sax\/ (see \S\ref{outburst})

Due to the sparse data for \he, we augmented the sample with a pair of bursts observed by \xte/PCA on 2009 June 12, separated by only 1.89~hr. The nearest observations in time preceeding (following) that observation were 13~d before (3~d after), so 
any possible test for periodicity over that time range would likely be uninformative given the likely phase drift of the burst sequence.
Assuming that the separation of 1.89~hr does reflect a regular recurrence time, this pair
 would be the most frequent bursts ever observed from this system \cite[cf. with][]{clark77b}.

The constraints on recurrence times for the superbursts from \su\/ is poor due to the low duty cycle of observations between the detected events. The superburst in our sample was preceded by another $1.75$ years prior. However, the inferred accretion column is approximately twice the ignition column, suggesting that the actual recurrence time is closer to $0.9$ years \citep{keek15a}. In addition to the properties of the burst itself reported in Table \ref{tab:bursts}, we constrained the ``quench time'' corresponding to the period following the superburst during which normal thermonuclear burning is suppressed. The first thermonuclear (H/He) event detected from \su\ following the superburst was on 
MJD 51985.5 ({\it BeppoSAX}\/ observation 10898), 
giving $t_{\rm quench}\leq22.8$~d. The observation during which that event was detected commenced 17.5~hr before the burst was detected, and had a duty cycle of $\approx70$\%, with interruptions due to the low-Earth orbit. The coverages suggests a $\approx30$\% chance that an earlier burst could have been missed, so that the quench time could have been shorter by up to 17.5~hr. 
However, no intervening observations were made to rule out additional bursts missed earlier than that observation.

\subsection{Bolometric corrections}
\label{bolcorr}

We measured the persistent flux in the 3--25~keV band from \xte\/ and {\it BeppoSAX}\/ observations by fitting commonly-used phenomenological models, and converted the integrated flux to a bolometric value by applying a bolometric correction for each observation. The correction was estimated by fitting to {\it RXTE}\/ PCA and HEXTE data  a broad-band model typically comprising an absorbed Comptonisation component \cite[{\tt compTT} in {\sc xspec};][]{tit94}, usually with a Gaussian  component representing Fe K$\alpha$ emission in the range 6.4--6.7~keV.
We measured the 3--25~keV flux from each fit, and then created an ideal response
{ (using {\tt dummyrsp} in {\sc xspec}) }
covering the energy range 0.1--1000~keV, and integrated the flux over this range, excluding the effects of neutral absorption. We estimated the bolometric correction as the ratio of the unabsorbed 0.1--1000~keV flux to the absorbed 3--25~keV flux; for the observations here, the correction was typically in the range 1.4--2.

This approach is reliable provided that the contribution to the bolometric flux outside the range to which PCA and HEXTE are sensitive (typically 3--100~keV) is small (or at least consistent). However, it is known that for \gs, that additional low-energy contributions may arise at times,
so that the persistent flux measured by \xte\/ in the 3--25~keV band is not always a reliable estimator of the accretion rate \cite[e.g.][]{thompson08}. In some observations, there is evidence for unusually low $\alpha$-values, suggesting that a significant fraction of the X-ray flux is emitted outside the \xte\/ band; contemporaneous {\it Chandra}\/ or {\it XMM-Newton}\/ observations suggest that this flux may be emitted as a low-temperature ($\lesssim1$~keV) thermal component. 
Thus, we avoided from our data selection bursts from such epochs identified by \cite{thompson08}, including those from 2003 April, and also the observations from 1998 June and 1997 November. 
An additional epoch of unusually low $\alpha\approx26$ is from observations in 2006 August, and was also excluded. The remaining set of three burst trains have broadly consistent $\alpha$-values, but we caution that there may yet be additional systematic errors contributing to the determination of the accretion rate in this (and possibly other) systems.

\subsection{Lightcurve fitting}
\label{lcfit}

Here we describe the approach by which the values of the neutron star radius, redshift, and surface gravity for \gs\ were chosen.
We 
replicated the analysis of \cite{heger07b} with updated analysis results and an improved method to determine the best-fit system parameters. We selected the train of bursts observed in 2000~September, with a recurrence time of 
4.177~hr, similar to that used previously. The 7 bursts observed with \xte\/ are augmented by the observation of 25 additional bursts with {\it BeppoSAX}\/ over the same interval, with the complete train spanning 12~d. 
However, the entire sequence of bursts is not consistent with a steady recurrence time, likely due to small variations in accretion rate. Thus, we restricted the sequence to a shorter interval covering the \xte\/ bursts, beginning on MJD~51810 and spanning almost 4~d (as plotted in Fig. \ref{example}).
We then compared the averaged lightcurve (calculated from the \xte\/ bursts only) with examples predicted by the {\sc kepler} code, tabulated in \cite{lampe16}.

\cite{heger07b} found excellent agreement between the observed lightcurve and a model referred to as ``A3'', with solar composition, an accretion rate of $1.58\times10^{-9}\ M_\odot\,{\rm yr}^{-1}$. In the sample of \cite{lampe16}, the accretion rate is quantified in terms of the Eddington value, defined as $1.75\times10^{-8}\ M_\odot\,{\rm yr}^{-1}$. Thus, for model A3, the accretion rate was $\dot{M}=0.09\dot{M}_{\rm Edd}$, and this model corresponds to ``a05d'' in the \cite{lampe16} sample. 

In our revised comparison, we simultaneously matched the lightcurve and the recurrence time, using a Markov-Chain Monte Carlo (MCMC) code {\tt emcee} \cite[]{emcee13} to marginalise over the three parameters of interest: the redshift $1+z$ (by which the lightcurve will be stretched when transforming into the observer's frame), the relative intensity scale, including the distance and redshift:
\begin{equation}
    F_X = \frac{L_X}{4\pi d^2\xi_b(1+z)}
\end{equation}
where $L_X$ is the burst luminosity predicted by {\sc kepler}, $d$ is the distance, $\xi_b$ 
{ takes into account the possible anisotropy of the burst emission (see also appendix \ref{distance},}
and $F_X$ the observed burst flux. A third parameter is related to the offset in time between the model and observation, and has no physical meaning.

Our initial runs with this code did not match the recurrence time, and to ensure that the model-predicted burst interval (when scaled by $1+z$) matched that observed, we weighted this parameter in the likelihood function by a factor of 50. This ensured in the subsequent comparisons that the redshifted model recurrence time matched that observed to within a few percent.

The comparison with  model {\tt a05d}, as used by \cite{heger07b}, gave good agreement with the observed lightcurves, although requiring a redshift of $1+z=1.392$. 
Following the approach of \cite{lampe16}, we transformed the model results (calculated in a Newtonian frame) to the observer's frame assuming that the gravity, and the Newtonian and the general-relativistic neutron star masses are identical. This implies that the radius required for the Newtonian and general-relativistic gravity will be different, and the corresponding radius (for a neutron star mass of $1.4\ M_\odot$) implied by our comparison of model {\tt a05d} was 8.5~km, well below the lower limit of 10.4~km inferred for a wide range of neutron stars \cite[e.g.][]{slb13}.

\begin{figure}
\begin{center}
\includegraphics[width=\columnwidth]{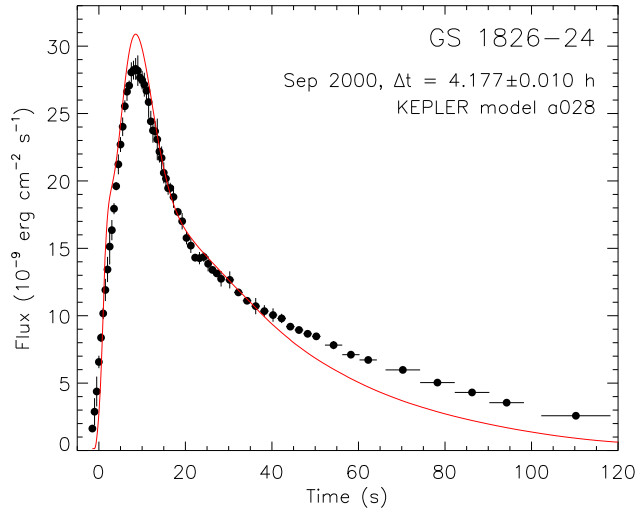}
\caption{Best-fitting comparison of averaged, observed burst lightcurves from \gs, with {\sc kepler} model {\tt a028}. The  symbols with errors show the average lightcurve observed by \xte\/ in 2000 September, at which time the recurrence time was 
4.177~hr
The model curve, rescaled on both axis based on the best-fit distance and redshift, is the red solid curve. The best-fitting distance and redshift, determined from an MCMC analysis, were 
$d\xi^{1/2}=\gsdist$~kpc and $1+z=\gsopz$, 
respectively.
}\label{fig:gslcompare}
\end{center}
\end{figure}

Thus, we made a limited exploration of alternative models, focussing on those with solar composition. The comparison with model {\tt a028} gave a best-fit redshift of 
$1.234$,
implying a radius of 
12.1~km,
well within the expected range. 
The accretion rate (in units of the Eddington rate) for this model is 0.082, which is somewhat higher than that inferred for the 2000~Sep epoch, at 0.0692; but may be consistent taking into account the persistent emission anistropy. \cite{heger07b} estimated the ratio of persistent to burst anisotropies at $\xi_p/\xi_b=1.55$, implying $\xi_p>1$ \cite[e.g.][]{fuji88}; in that case, the luminosity (and hence accretion rate) inferred from the persistent flux would be higher, bringing the observation estimate roughly in line with the value assumed for run {\tt a028}.
The corresponding distance is $d\xi_b^{1/2} = \gsdist$~kpc. 
We do not quote errors on these parameters, as the lightcurve comparison (Fig. \ref{fig:gslcompare}) is not of sufficient quality to have confidence in the posterior distributions 
{ (the estimated probability densities for the model parameters)}. However, the comparison is of comparable quality as in \cite{heger07b}, and so for the present purposes, we adopt the values for the redshift and distance.

\subsection{Outburst fitting}
\label{outburst}

The approach adopted for \sax\/ was necessarily different, as this source
has not exhibited regular bursts in previous observations. The persistent flux
(and hence accretion rate) varied significantly over the time at which the
bursts were observed in 2002 October, and so the burst separations (and hence, presumably,
the recurrence times) vary significantly. Furthermore, the burst separations are all longer than 
{ 12~hr},
so that many data gaps are present between each burst, introduced by the interruptions in the visibility of the source due to the observing satellite's $\approx90$-min low-Earth orbit.

\cite{gal06c} compared the observed burst times with the predictions of a numerical ignition model, to constrain the system parameters. 
For the purposes of this paper, we have revised the analysis of
those authors,
to reconcile the changes in the \xte/PCA calibration
since that work was completed, to correctly treat the persistent and burst anisotropy factors, and to use a more rigorous parameter space exploration to determine the best-fit system parameters and uncertainties. 

We matched the burst times, fluences and $\alpha$-measurements to the {\sc settle} code developed by \cite{cb00} using 
{ MCMC and marginalising over }
the accreted hydrogen fraction $X_0$, CNO metallicity $Z_{\rm CNO}$, base flux $Q_b$, and scaling factors 
incorporating the source distance and emission anisotropy.
The {\sc settle} model predicts burst parameters in the neutron star frame, given an inferred accretion rate. Necessarily, we adopt the average inferred accretion rate (based on linear interpolation between each observation) for each burst interval, as {\sc settle} does not simulate varying $\dot{m}$.
The redshift $1+z$ and (Newtonian) surface gravity $g$ are fixed in the {\sc settle} code, at the values listed in Table \ref{tab:sources} so the results are implicitly for those values.

The best-fit parameters we report are dependent on the fidelity of the 
model, which 
does not take into account all of the underlying physics, such as the effects of thermal and compositional inertia. More detailed models, such as {\sc kepler} \cite[]{woos04} can provide more rigorous modelling of the underlying physics, however are computationally expensive and so impractical for the Monte Carlo approach adopted here.
However, 
{\sc kepler} simulations to reproduce the bursts observed from \sax\ during the 2002 outburst have been performed recently,  demonstrating that the assumption of a constant accretion rate between bursts results in the recurrence time being underestimated, when compared to the scenario in which the accretion rate decreases between bursts (Johnston et al. 2017, in preparation). 
To account for this bias, we corrected the recurrence time predictions of {\sc settle} using correction factors determined by {\sc kepler}. The correction factor depends on the $\dot{m}$ gradient, and thus is different for each burst. We adopt correction factors for each of the seven predicted bursts of \sax, which vary from $\approx$ 1 for the first burst to $\approx$ 1.2 for the final burst. 

We infer the persistent and burst anisotropy factors ($\xi_p$ and $\xi_b$ respectively) from the scaling factors and find $\xi_p$ = 1.05$\pm$0.01 and $\xi_b$ = 0.91$^{+0.04}_{-0.03}$. This implies an inclination of 62$\pm$1$\degree$. Note that the error on the inclination is statistical only, and does not include the contribution from uncertainties in the mass and radius, due to the fixed values adopted in the {\sc settle} model.

As with the  { analysis of \cite{gal06c}}, our maximum-likelihood solution predicts an additional pair of bursts between the first pair of observed bursts, on MJD~52562.41363 and 52564.30515 \cite[\#1 and 2 from][]{bcatalog}.
These events would have fallen in data gaps, which explains why they were not observed.
We thus report the burst
separations for the last two bursts observed during the 2002 October
outburst as the recurrence time; for the second burst, we estimate the recurrence time inferred
from the burst train modelling based on the simulated time of the preceding event. 
{ We note that the inferred recurrence time of 16.55~hr for the second burst observed with \xte\/ is comfortably within the overall range for the source, with the minimum separation of 12.6~hr set by a pair of bursts observed by {\it BeppoSAX}\/ \cite[]{zand98c}. }

\section{RESULTS}
\label{results}

\begin{figure}
\begin{center}
\vspace{-0.7cm}
\includegraphics[width=\columnwidth]{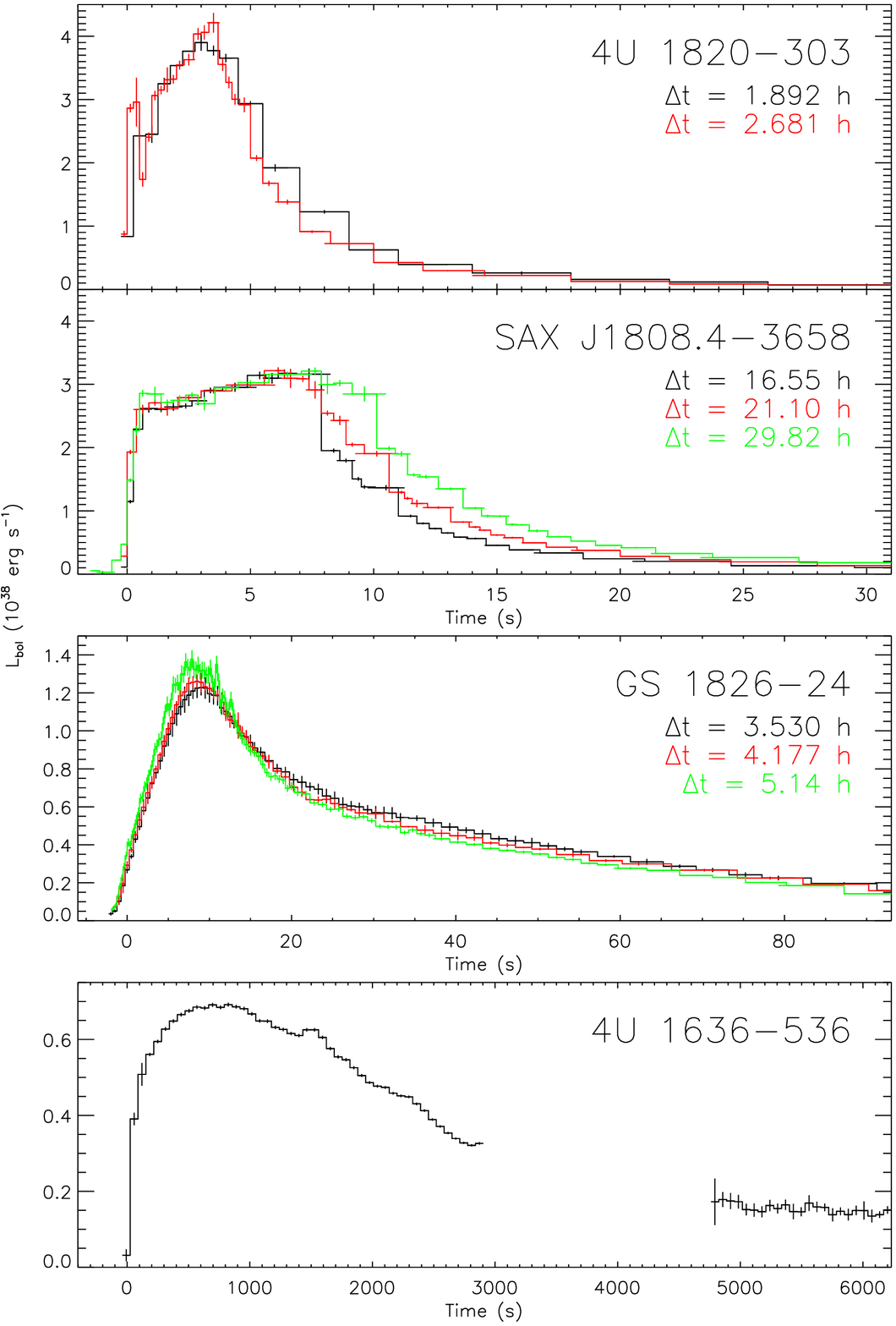}
\caption{Representative burst lightcurves from \he, \sax, \gs\  and \su\ observed by
the {\it Rossi X-ray Timing Explorer}. In each panel we plot the 
lightcurve at the indicated recurrence time, with flux converted to
luminosity at the distances indicated in Table \ref{tab:sources}.
The bursts contributing to each profile are listed in Table \ref{tab:bursts};
where more than one profile is observed by \xte\/ in the train, we show
the average profile.
Note the change in $x$- and $y$-axis between the top panels (\he\ and \sax) and the
lower two panels.
}\label{fig:examples}
\end{center}
\end{figure}

The derived (or adopted) properties for the sources of interest are listed in Table \ref{tab:sources}.
The properties of the bursts selected for our sample are summarised in Table \ref{tab:bursts}.
We plot the burst lightcurves for each of the sequences in Figure \ref{fig:examples}. 
The accompanying material\footnote{The lightcurves are intended to be made available in the PASA data store upon publication of the paper, but in the meantime can be found at \url{http://burst.sci.monash.edu/reference}}. for this paper includes lightcurves (averaged in cases where there are more than one \xte\/ burst in the sequence) for comparison with numerical model results.

The burst lightcurves show clear differences related to the inferred composition and accretion rate. For \he, the burst lightcurves are short, with durations (defined as the interval over which the luminosity exceeds 10\% of the maximum) of 15~s or so. Each of these bursts exhibits strong { photospheric} radius expansion, and the quality of the spectral fits during the radius expansion episodes are relatively poor (reduced-$\chi^2$ values in the range 2--8). As $\dot{m}$ increases, the burst timescale increases slightly ($\tau=6.24$ to $6.55$~s), which qualitatively supports the presence of a small amount of H in the accreted fuel. With smaller recurrence time, there is less time to burn the accreted H, and the mass fraction at ignition will be larger.

The bursts from \sax, although also thought to be powered primarily by He, have a notably different morphology, with an extended radius-expansion phase at roughly constant luminosity.
The timescale variation for \sax\/ is in the opposite direction, with the burst timescale and duration becoming shorter at higher accretion rates. Here the recurrence time is long enough that any accreted H has already been exhausted at the base, so that lower accretion rate allows a larger pile of He to accumulate. 
{ The resulting increased fluence thus requires a longer interval emitting at the Eddington luminosity,  with some of the additional energy likely exported as kinetic energy of outflowing material. }

The bursts from \gs\/ are different again, showing the much slower ($\approx5$~s) rises and long rp-process powered tails characteristic of these mixed H/He bursts. The peak luminosity is significantly smaller than for the He-rich bursts, and durations are approximately a factor of 5 longer.
The burst profiles show only very subtle variations from epoch to epoch, with the peak flux decreasing slightly, while the timescale increases, as the accretion rate increases.

The superburst profile is principally distinguished from the shorter, H or H/He events by the orders-of-magnitude longer timescale, with a duration of more than 2~hr.

\begin{table*}[h]
  \begin{sideways}
  \begin{minipage}{\textheight} 

\begin{center}
\caption{Properties of thermonuclear bursts observed from target sources by
the {\it Rossi X-ray Timing Explorer}}\label{tab:bursts}
\begin{tabular}{lccccccccc}
& $N_{\rm burst}$ & burst & $\Delta t$ & $F_{\rm per}^b$ & 
& $\dot{m}$ & $E_b$ & $F_{\rm pk}$ &
\\
Epoch & ($N_{\rm av}$) & IDs$^a$ & (hr) 
& ($10^{-9}\ \epcs$) & $c_{\rm bol}$  & ($\dot{m}_{\rm Edd}$) & 
  ($10^{-6}\ \epc$) & ($10^{-9}\ \epcs$) & $\alpha$ \\
\hline
\multicolumn{10}{c}{\gs} \\
\hline
1998 Jun & 6(1) & 7 & $5.14\pm0.07$ & $1.167\pm0.006$ & $1.806\pm0.009$ & 0.0513 & $1.102\pm0.011$ & $30.9\pm1.0$ & $34.2\pm0.5$ \\ 
2000 Sep & 11(7) & 14--20 & $4.177\pm0.010$ & $1.593\pm0.017$ & $1.787\pm0.003$ & 0.0692 & $1.126\pm0.016$ & $29.1\pm0.5$ & $38.6\pm0.3$ \\ 
2007 Mar & 10(7) & 58--65 & $3.530\pm0.004$ & $1.87\pm0.02$ & $1.751\pm0.003$ & 0.0796 & $1.18\pm0.04$ & $28.4\pm0.4$ & $35.3\pm1.0$ \\ 
\hline
\multicolumn{10}{c}{\sax} \\

\hline
2002 Oct & 1(1) & 2 & \saxdt$^c$ & 2.541--2.298 & $2.085\pm 0.019$ & 0.0472$^d$ & $2.649\pm0.018$ & $229\pm4$ & $114.4\pm1.9$ \\ 
         & 1(1) & 3 & 21.10 & 2.298--1.946 & $2.13\pm0.04$ & 0.0432$^d$ & $2.990\pm0.017$ & $232\pm4$ & $118.2\pm1.9$ \\ 
         & 1(1) & 4 & 29.82 & 1.946--1.826 & $2.157\pm0.002$ & 0.0384$^d$ & $3.46\pm0.02$ & $232\pm4$ & $128.2\pm2.1$ \\ 
\hline
\multicolumn{10}{c}{\he} \\
\hline
1997 May 4 & 3(1) & 1 & $2.681\pm0.007$ & $3.72\pm0.18$ & $1.45\pm0.09$ & 0.144 & $0.381\pm0.003$ & $61\pm2$ & $138.5\pm1.4$ \\
2009 Jun 12 & 2(2) & & $1.892^e$ & $5.70\pm0.04$ & 1.4981 & 0.226 & $0.371\pm0.010$ & $56.6\pm1.4$ & $160.1\pm1.8$\\ 
\hline
\multicolumn{10}{c}{\su} \\
\hline
2001 Feb & 1(1) & & $\leq 1.53\times10^{4}$~$^f$ & $4.73\pm0.03$ & 1.5346 & 0.167 & $110\pm9$~$^g$ & $21.9\pm0.6$~$^g$ \\
\hline
\end{tabular}
\medskip\\
\end{center}
$^a$ burst index number in the catalog of \cite{bcatalog}.\\
$^b$ persistent flux in the energy range 3--25 keV \\
$^c$ inferred from ignition model comparisons \\
$^d$ calculated based on linear interpolation between observations falling within the burst interval \\
$^e$ only two bursts detected, so may not have been part of a regular train \\
$^f$ recurrence time upper limit of 1.75~yr based on the next earliest event, detected by the \xte/ASM \cite[]{kuul04} \\
$^g$ values taken from the ``optimal'' fits of \cite{keek14a}
 \end{minipage}
 \end{sideways}
\end{table*}

\section{DISCUSSION}
\label{discussion}

We have assembled a  sample of burst observations intended for comparisons with, and between, numerical models. Comparisons of different model codes for the same test cases will allow estimates of the intrinsic uncertainty that affects burst simulations. These uncertainties may also depend on the particular type of burst being simulated; 
{ that is, the variation in model predictions for some types of bursts may
be larger than for others. } For example, \cite{gal06c} suggest that the
simple analytic model adopted for the simulations of bursts from \sax\
offer { comparable fidelity to} more sophisticated models, { including
{\sc kepler}. This is less true } for simulations of H-rich bursts from \gs\ observed at higher accretion rates, { as the simple models do not include the effects of compositional or thermal inertia.} 

It is also anticipated that detailed comparisons of the burst measurements here may allow more stringent constraints to be placed on the neutron star parameters, as has been demonstrated as a proof-of-principle in sections \S\ref{lcfit} and \S\ref{outburst}.

The anticipated procedure to replicate each of the burst measurements in Table \ref{tab:bursts} is as follows

\begin{enumerate}
\item Set the surface gravity $g$ and neutron star radius $R$ to the values appropriate for the source of interest in Table \ref{tab:sources}
\item Set the composition in the accreted material, $X_0$ and $Z_{\rm CNO}$
\item Perform a model run with a constant accretion rate chosen from one of the entries for that source from Table \ref{tab:bursts}, and extract the burst luminosity as a function of time. 
\item Re-scale the predicted luminosity appropriately for the redshift parameter $1+z$ for this source, listed in Table \ref{sources}, and also the burst anisotropy parameter $\xi_b$, where known. For the second and third bursts from \sax, the 
predicted recurrence time and burst fluence
may be adjusted by the factor described in \S\ref{outburst} to account for the effects of the declining accretion rate during each burst
\item Calculate from the simulated burst train the recurrence time $\Delta t_{\rm pred}$, and from the burst lightcurve, the burst fluence, peak flux, $\alpha$-value, and timescale
\item Compare these predictions with the observed lightcurve and the measured values in Table \ref{tab:bursts}.  
\end{enumerate}

In the event that the model recurrence time is significantly longer (shorter) than the observation, the cause may be that the true accretion rate is different from the estimate due to the unknown degree of anisotropy $\xi_p$ of the persistent emission. In that case, the adopted accretion rate might be increased (decreased), by roughly the ratio of the predicted and observed recurrence times, and the procedure repeated from step 3 onwards.

One additional parameter that is commonly included in burst simulations is the heat flux from below the simulation region, commonly parameterised as a ``base flux'' $Q_b$. This quantity is not well constrained by observations, and furthermore depends on the accretion rate \cite[e.g. Fig. 18 of ][]{cumming06}; most studies to date have adopted values in the range 0.1--0.15~MeV~nucleon$^{-1}$. For the purposes of model-to-model comparisons we recommend adopting a value of 0.1, but suggest that for the purposes of matching the observations, this parameter can be considered free within the range 0.1--0.8~MeV~nucleon$^{-1}$, depending upon the choice of core neutrino emissivity model.

The  approach to replicate the superburst observation is necessarily different.
Modeling a superburst by accreting hydrogen/helium comes at great computational expense, because hundreds of hydrogen/helium flashes must be simulated to create the carbon fuel. Furthermore, it remains challenging to reproduce the observationally inferred carbon mass fraction \citep[e.g.,][]{stevens14,keek16}. An alternative approach is to directly accrete a mixture of carbon and heavier elements \citep[e.g., Fe or Ru;][]{cumming01,schatz03}. For the 2001 superburst from 4U~1636$-$536, we recommend a fuel composition of $26\%$ $^{12}$C and $74\%$ $^{56}$Fe \citep{keek15a}. Such simulations are able to describe the observed superburst light curves in detail, but generally predict larger ignition depths than observed \citep{keek11}.

The ultimate goal of this study is 
to constrain the rates of particular nuclear reactions, or masses of particular isotopes, based on the burst observations. It has been shown that several reactions are likely to affect the shape of the burst lightcurve \cite[e.g.][]{cyburt16}. Similarly, nuclear masses have also been shown to have an effect \cite[]{so16}. It is hoped that the carefully-assembled and calibrated data presented here offer the first step on the road towards precision nuclear physics from thermonuclear bursts.

\begin{acknowledgements}
This work benefited from discussions at the International Symposium on Neutron Stars in the Multi-Messenger Era in May 2016, and more broadly by support from the National Science Foundation under Grant No. PHY-1430152 (JINA Center for the Evolution of the Elements).
The authors are grateful for support received as part of the International Team on Nuclear Reactions in Superdense Matter by the International Space Science Institute in Bern, Switzerland.
The data analysed here are derived from preliminary analysis results of the Multi-INstrument Burst ARchive (MINBAR), which is supported under the Australian Academy of Science's Scientific Visits to Europe program, and the Australian Research Council's Discovery Projects and Future Fellowship funding schemes.
LK is supported by NASA under award number NNG06EO90A.
\end{acknowledgements}

\appendix

\section{Source distances}
\label{distance}

An important source of systematic uncertainty in the determination of persistent and burst luminosities is the source distance $d$, and in combination, the degree of  anisotropy of the burst and persistent emission, parameterised as $\xi_b$ and $\xi_p$, respectively.
Anisotropy of the burst (and persistent) flux is expected given the strongly non-spherically symmetric geometry introduced by the accretion disk and binary companion. 
While the distances to burst sources have been estimated with moderate precision from observations of radius-expansion bursts \cite[e.g.][]{bcatalog}, these measurements  implicitly also depend on the burst anisotropy.
Strictly speaking, distances estimated in such a manner correspond to $d\xi_b^{1/2}$, 
and in general it is not possible to separate these quantities to derive the physical distance $d$, independently.

The definition of the anisotropy factor $\xi_b$ follows that of \cite[]{fuji88}, such that for a total burst luminosity $L_b$, the measured burst flux would be given by 
\begin{equation}
F_b = \frac{L_b}{4\pi d^2\xi_b}
\end{equation}
From the above relation, we understand that $\xi_b=1$ corresponds to isotropic emission in all directions, while $\xi_b<1$ indicates that the burst flux toward the observer is {\em in excess} of the isotropic value, i.e. preferential beaming in our direction. 
As pointed out by \cite{he16}, the values of $\xi_b$ and $\xi_p$ depend upon the system inclination and the structure of the accretion disk, neither of which are well constrained.

For \su, 
we estimated the distance based on measurements of the burst emission from photospheric radius-expansion (PRE) bursts in MINBAR, assuming that they reach the Eddington limit.
This follows the approach of \cite{bcatalog}, but the burst flux measurements have been updated with the latest PCA response matrices, and the effects of deadtime in the burst spectra have been included. The overall effect on the distances (compared to the previous analyses) is a reduction of 5--10\%.
We also incorporated the assumed gravitational redshift at the surface, as listed in Table \ref{tab:sources}.

The anisotropy of burst emission for \su{} can be estimated from the inclination of the binary system with respect to the observer's line of sight \citep[e.g.,][]{fuji88}.  Optical observations constrain the inclination angle to the range $36^\circ-74^\circ$ \citep{casares06}, which implies $0.7\lesssim \xi_\mathrm{b}^{-1}\lesssim 1.3$ for a thin disk \citep[larger values for smaller angles;][]{he16}. The observation of a broad fluorescent iron line favors a large inclination angle \citep{pandel08}, but this depends strongly on the interpretation of the reflection spectrum. Similarly, different constraints can be derived from the reflection signal during the superburst from this source \citep{keek15a}. Therefore, an uncertainty of several tens of percents must be taken into account for the burst luminosity of \su{}.

For \gs\ and \sax, we instead adopted distances derived from comparing the burst measurements with models, as described in sections \S\ref{lcfit} and \S\ref{outburst}.

\cite{zamfir12a} derived an upper limit on the distance  (combined with the burst anisotropy) for \gs\ of $d\xi_b^{1/2}<5.5$~kpc.
While our lightcurve comparisons for this source do not take into account the blackbody normalisation, as was analysed by \cite{zamfir12a}, our derived system constraints are otherwise broadly consistent with those authors. However, our best-fit value based on the comparison with model {\tt a028} of \cite{lampe16} is \gsdist~kpc, slightly higher than the inferred limit. 

One additional consistency check that we can make is to compare the expected Eddington flux for the adopted system parameters, with that measured from the first photospheric radius-expansion burst from the source, observed in 2014 June with {\it NuSTAR}\/ \cite[]{chenevez16}. Assuming that the degree of burst anisotropy $\xi_b$ does not change during radius expansion, and that the peak flux is achieved at touchdown (when the expanding photosphere has returned to the neutron star surface), the expected flux would be \cite[e.g.][]{bcatalog} $3.7\times10^{-8}\ \epcs$. This is well within the confidence interval of $(4.0\pm0.3)\times10^{-8}\ \epcs$ derived for the radius-expansion burst observed by \cite{chenevez16}.
In the absence of a more detailed comparison with {\sc kepler} simulations, which would allow us to incorporate the results from all three burst trains presented in Table \ref{tab:bursts},  we adopt the current value, but note that the true value may be lower.

For \he, we adopted the distance to the host globular cluster, NGC 6624 
\cite[]{kuul03a}.


\begin{thebibliography}{65}
\expandafter\ifx\csname natexlab\endcsname\relax\def\natexlab#1{#1}\fi

\bibitem[{{Augusteijn} {et~al.}(1998){Augusteijn}, {van der Hooft}, {de Jong},
  {van Kerkwijk}, \& {van Paradijs}}]{august98}
{Augusteijn}, T., {van der Hooft}, F., {de Jong}, J.~A., {van Kerkwijk}, M.~H.,
  \& {van Paradijs}, J. 1998, \aap, 332, 561

\bibitem[{{Casares} {et~al.}(2006){Casares}, {Cornelisse}, {Steeghs},
  {Charles}, {Hynes}, {O'Brien}, \& {Strohmayer}}]{casares06}
{Casares}, J., {Cornelisse}, R., {Steeghs}, D., {Charles}, P.~A., {Hynes},
  R.~I., {O'Brien}, K., \& {Strohmayer}, T.~E. 2006, \mnras, 373, 1235

\bibitem[{{Chakrabarty} \& {Morgan}(1998)}]{chak98d}
{Chakrabarty}, D. \& {Morgan}, E.~H. 1998, \nat, 394, 346

\bibitem[{{Chakrabarty} {et~al.}(2003){Chakrabarty}, {Morgan}, {Muno},
  {Galloway}, {Wijnands}, {van der Klis}, \& {Markwardt}}]{chak03a}
{Chakrabarty}, D., {Morgan}, E.~H., {Muno}, M.~P., {Galloway}, D.~K.,
  {Wijnands}, R., {van der Klis}, M., \& {Markwardt}, C.~B. 2003, \nat, 424, 42

\bibitem[{{Chenevez} {et~al.}(2016){Chenevez}, {Galloway}, {in 't Zand},
  {Tomsick}, {Barret}, {Chakrabarty}, {F{\"u}rst}, {Boggs}, {Christensen},
  {Craig}, {Hailey}, {Harrison}, {Romano}, {Stern}, \& {Zhang}}]{chenevez16}
{Chenevez}, J., {Galloway}, D.~K., {in 't Zand}, J.~J.~M., {Tomsick}, J.~A.,
  {Barret}, D., {Chakrabarty}, D., {F{\"u}rst}, F., {Boggs}, S.~E.,
  {Christensen}, F.~E., {Craig}, W.~W., {Hailey}, C.~J., {Harrison}, F.~A.,
  {Romano}, P., {Stern}, D., \& {Zhang}, W.~W. 2016, \apj, 818, 135

\bibitem[{{Clark} {et~al.}(1977){Clark}, {Li}, {Canizares}, {Hayakawa},
  {Jernigan}, \& {Lewin}}]{clark77b}
{Clark}, G.~W., {Li}, F.~K., {Canizares}, C., {Hayakawa}, S., {Jernigan}, G.,
  \& {Lewin}, W.~H.~G. 1977, \mnras, 179, 651

\bibitem[{{Cornelisse} {et~al.}(2003){Cornelisse}, {in 't Zand}, {Verbunt},
  {Kuulkers}, {Heise}, {den Hartog}, {Cocchi}, {Natalucci}, {Bazzano}, \&
  {Ubertini}}]{corn03a}
{Cornelisse}, R., {in 't Zand}, J.~J.~M., {Verbunt}, F., {Kuulkers}, E.,
  {Heise}, J., {den Hartog}, P.~R., {Cocchi}, M., {Natalucci}, L., {Bazzano},
  A., \& {Ubertini}, P. 2003, \aap, 405, 1033

\bibitem[{{Cumming}(2003)}]{cumming03}
{Cumming}, A. 2003, \apj, 595, 1077

\bibitem[{{Cumming} \& {Bildsten}(2000)}]{cb00}
{Cumming}, A. \& {Bildsten}, L. 2000, \apj, 544, 453

\bibitem[{{Cumming} \& {Bildsten}(2001)}]{cumming01}
---. 2001, \apjl, 559, L127

\bibitem[{{Cumming} {et~al.}(2006){Cumming}, {Macbeth}, {Zand}, \&
  {Page}}]{cumming06}
{Cumming}, A., {Macbeth}, J., {Zand}, J.~J.~M.~i., \& {Page}, D. 2006, \apj,
  646, 429

\bibitem[{{Cyburt} {et~al.}(2016){Cyburt}, {Amthor}, {Heger}, {Johnson},
  {Keek}, {Meisel}, {Schatz}, \& {Smith}}]{cyburt16}
{Cyburt}, R.~H., {Amthor}, A.~M., {Heger}, A., {Johnson}, E., {Keek}, L.,
  {Meisel}, Z., {Schatz}, H., \& {Smith}, K. 2016, \apj, 830, 55

\bibitem[{{Fisker} {et~al.}(2006){Fisker}, {G{\"o}rres}, {Wiescher}, \&
  {Davids}}]{fisker06}
{Fisker}, J.~L., {G{\"o}rres}, J., {Wiescher}, M., \& {Davids}, B. 2006, \apj,
  650, 332

\bibitem[{Foreman-Mackey {et~al.}(2013)Foreman-Mackey, Hogg, Lang, \&
  Goodman}]{emcee13}
Foreman-Mackey, D., Hogg, D.~W., Lang, D., \& Goodman, J. 2013, Publications of
  the Astronomical Society of the Pacific, 125, pp. 306

\bibitem[{{Fujimoto}(1988)}]{fuji88}
{Fujimoto}, M.~Y. 1988, \apj, 324, 995

\bibitem[{{Fujimoto} {et~al.}(1981){Fujimoto}, {Hanawa}, \& {Miyaji}}]{fhm81}
{Fujimoto}, M.~Y., {Hanawa}, T., \& {Miyaji}, S. 1981, \apj, 247, 267

\bibitem[{{Galloway} \& {Cumming}(2006)}]{gal06c}
{Galloway}, D.~K. \& {Cumming}, A. 2006, \apj, 652, 559

\bibitem[{{Galloway} {et~al.}(2004){Galloway}, {Cumming}, {Kuulkers},
  {Bildsten}, {Chakrabarty}, \& {Rothschild}}]{gal03d}
{Galloway}, D.~K., {Cumming}, A., {Kuulkers}, E., {Bildsten}, L.,
  {Chakrabarty}, D., \& {Rothschild}, R.~E. 2004, \apj, 601, 466

\bibitem[{{Galloway} {et~al.}(2008){Galloway}, {Muno}, {Hartman}, {Psaltis}, \&
  {Chakrabarty}}]{bcatalog}
{Galloway}, D.~K., {Muno}, M.~P., {Hartman}, J.~M., {Psaltis}, D., \&
  {Chakrabarty}, D. 2008, \apjs, 179, 360

\bibitem[{{Grindlay} {et~al.}(1976){Grindlay}, {Gursky}, {Schnopper},
  {Parsignault}, {Heise}, {Brinkman}, \& {Schrijver}}]{grindlay76}
{Grindlay}, J., {Gursky}, H., {Schnopper}, H., {Parsignault}, D.~R., {Heise},
  J., {Brinkman}, A.~C., \& {Schrijver}, J. 1976, \apjl, 205, L127

\bibitem[{{He} \& {Keek}(2016)}]{he16}
{He}, C.-C. \& {Keek}, L. 2016, \apj, 819, 47

\bibitem[{{Heger} {et~al.}(2007){Heger}, {Cumming}, {Galloway}, \&
  {Woosley}}]{heger07b}
{Heger}, A., {Cumming}, A., {Galloway}, D.~K., \& {Woosley}, S.~E. 2007, \apjl,
  671, L141

\bibitem[{{in 't Zand} {et~al.}(1998){in 't Zand}, {Heise}, {Muller},
  {Bazzano}, {Cocchi}, {Natalucci}, \& {Ubertini}}]{zand98c}
{in 't Zand}, J.~J.~M., {Heise}, J., {Muller}, J.~M., {Bazzano}, A., {Cocchi},
  M., {Natalucci}, L., \& {Ubertini}, P. 1998, \aap, 331, L25

\bibitem[{{in 't Zand} {et~al.}(2004){in 't Zand}, {Verbunt}, {Heise},
  {Bazzano}, {Cocchi}, {Cornelisse}, {Kuulkers}, {Natalucci}, \&
  {Ubertini}}]{zand04b}
{in 't Zand}, J.~J.~M., {Verbunt}, F., {Heise}, J., {Bazzano}, A., {Cocchi},
  M., {Cornelisse}, R., {Kuulkers}, E., {Natalucci}, L., \& {Ubertini}, P.
  2004, in Proceedings of the 2nd BeppoSAX Conference: "The Restless
  High-Energy Universe", Amsterdam, 5--9 May 2003, ed. E.~P.~J. {van den
  Heuvel}, R.~A.~M.~J. {Wijers}, \& J.~J.~M. {in 't Zand}, Vol. 132, 486--495

\bibitem[{{in't Zand} {et~al.}(2012){in't Zand}, {Homan}, {Keek}, \&
  {Palmer}}]{zand12a}
{in't Zand}, J.~J.~M., {Homan}, J., {Keek}, L., \& {Palmer}, D.~M. 2012, \aap,
  547, A47

\bibitem[{{Jager} {et~al.}(1997){Jager}, {Mels}, {Brinkman}, {Galama},
  {Goulooze}, {Heise}, {Lowes}, {Muller}, {Naber}, {Rook}, {Schuurhof},
  {Schuurmans}, \& {Wiersma}}]{jager97}
{Jager}, R., {Mels}, W.~A., {Brinkman}, A.~C., {Galama}, M.~Y., {Goulooze}, H.,
  {Heise}, J., {Lowes}, P., {Muller}, J.~M., {Naber}, A., {Rook}, A.,
  {Schuurhof}, R., {Schuurmans}, J.~J., \& {Wiersma}, G. 1997, \aaps, 125, 557

\bibitem[{{Jahoda} {et~al.}(1996){Jahoda}, {Swank}, {Giles}, {Stark},
  {Strohmayer}, {Zhang}, \& {Morgan}}]{xte96}
{Jahoda}, K., {Swank}, J.~H., {Giles}, A.~B., {Stark}, M.~J., {Strohmayer}, T.,
  {Zhang}, W., \& {Morgan}, E.~H. 1996, \procspie, 2808, 59

\bibitem[{{Keek} {et~al.}(2014){Keek}, {Ballantyne}, {Kuulkers}, \&
  {Strohmayer}}]{keek14a}
{Keek}, L., {Ballantyne}, D.~R., {Kuulkers}, E., \& {Strohmayer}, T.~E. 2014,
  \apj, 789, 121

\bibitem[{{Keek} {et~al.}(2015){Keek}, {Cumming}, {Wolf}, {Ballantyne},
  {Suleimanov}, {Kuulkers}, \& {Strohmayer}}]{keek15a}
{Keek}, L., {Cumming}, A., {Wolf}, Z., {Ballantyne}, D.~R., {Suleimanov},
  V.~F., {Kuulkers}, E., \& {Strohmayer}, T.~E. 2015, \mnras, 454, 3559

\bibitem[{{Keek} {et~al.}(2010){Keek}, {Galloway}, {in't Zand}, \&
  {Heger}}]{keek10}
{Keek}, L., {Galloway}, D.~K., {in't Zand}, J.~J.~M., \& {Heger}, A. 2010,
  \apj, 718, 292

\bibitem[{{Keek} \& {Heger}(2011)}]{keek11}
{Keek}, L. \& {Heger}, A. 2011, \apj, 743, 189

\bibitem[{{Keek} \& {Heger}(2016)}]{keek16}
---. 2016, \mnras, 456, L11

\bibitem[{{King} \& {Watson}(1986)}]{kw86}
{King}, A.~R. \& {Watson}, M.~G. 1986, \nat, 323, 105

\bibitem[{{Kuulkers}(2009)}]{kuul09b}
{Kuulkers}, E. 2009, The Astronomer's Telegram, 2140

\bibitem[{{Kuulkers} {et~al.}(2003){Kuulkers}, {den Hartog}, {in 't Zand},
  {Verbunt}, {Harris}, \& {Cocchi}}]{kuul03a}
{Kuulkers}, E., {den Hartog}, P.~R., {in 't Zand}, J.~J.~M., {Verbunt},
  F.~W.~M., {Harris}, W.~E., \& {Cocchi}, M. 2003, \aap, 399, 663

\bibitem[{{Kuulkers} {et~al.}(2004){Kuulkers}, {in 't Zand}, {Homan}, {van
  Straaten}, {Altamirano}, \& {van der Klis}}]{kuul04}
{Kuulkers}, E., {in 't Zand}, J., {Homan}, J., {van Straaten}, S.,
  {Altamirano}, D., \& {van der Klis}, M. 2004, in X-ray Timing 2003: Rossi and
  Beyond, ed. P.~{Kaaret}, F.~K. {Lamb}, \& J.~H. {Swank}, Vol. 714 (Melville,
  NY: AIP), 257--260

\bibitem[{{Lampe} {et~al.}(2016){Lampe}, {Heger}, \& {Galloway}}]{lampe16}
{Lampe}, N., {Heger}, A., \& {Galloway}, D.~K. 2016, \apj, 819, 46

\bibitem[{{Lewin} {et~al.}(1993){Lewin}, {van Paradijs}, \& {Taam}}]{lew93}
{Lewin}, W. H.~G., {van Paradijs}, J., \& {Taam}, R.~E. 1993, \ssr, 62, 223

\bibitem[{{Lund} {et~al.}(2003){Lund}, {Budtz-J{\o}rgensen}, {Westergaard},
  {Brandt}, {Rasmussen}, {Hornstrup}, {Oxborrow}, {Chenevez}, {Jensen},
  {Laursen}, {Andersen}, {Mogensen}, {Rasmussen}, {Om{\o}}, {Pedersen},
  {Polny}, {Andersson}, {Andersson}, {K{\"a}m{\"a}r{\"a}inen}, {Vilhu},
  {Huovelin}, {Maisala}, {Morawski}, {Juchnikowski}, {Costa}, {Feroci},
  {Rubini}, {Rapisarda}, {Morelli}, {Carassiti}, {Frontera}, {Pelliciari},
  {Loffredo}, {Mart{\'{\i}}nez N{\'u}{\~n}ez}, {Reglero}, {Velasco}, {Larsson},
  {Svensson}, {Zdziarski}, {Castro-Tirado}, {Attina}, {Goria}, {Giulianelli},
  {Cordero}, {Rezazad}, {Schmidt}, {Carli}, {Gomez}, {Jensen}, {Sarri},
  {Tiemon}, {Orr}, {Much}, {Kretschmar}, \& {Schnopper}}]{lund03}
{Lund}, N., {Budtz-J{\o}rgensen}, C., {Westergaard}, N.~J., {Brandt}, S.,
  {Rasmussen}, I.~L., {Hornstrup}, A., {Oxborrow}, C.~A., {Chenevez}, J.,
  {Jensen}, P.~A., {Laursen}, S., {Andersen}, K.~H., {Mogensen}, P.~B.,
  {Rasmussen}, I., {Om{\o}}, K., {Pedersen}, S.~M., {Polny}, J., {Andersson},
  H., {Andersson}, T., {K{\"a}m{\"a}r{\"a}inen}, V., {Vilhu}, O., {Huovelin},
  J., {Maisala}, S., {Morawski}, M., {Juchnikowski}, G., {Costa}, E., {Feroci},
  M., {Rubini}, A., {Rapisarda}, M., {Morelli}, E., {Carassiti}, V.,
  {Frontera}, F., {Pelliciari}, C., {Loffredo}, G., {Mart{\'{\i}}nez
  N{\'u}{\~n}ez}, S., {Reglero}, V., {Velasco}, T., {Larsson}, S., {Svensson},
  R., {Zdziarski}, A.~A., {Castro-Tirado}, A., {Attina}, P., {Goria}, M.,
  {Giulianelli}, G., {Cordero}, F., {Rezazad}, M., {Schmidt}, M., {Carli}, R.,
  {Gomez}, C., {Jensen}, P.~L., {Sarri}, G., {Tiemon}, A., {Orr}, A., {Much},
  R., {Kretschmar}, P., \& {Schnopper}, H.~W. 2003, \aap, 411, L231

\bibitem[{{Narayan} \& {Heyl}(2003)}]{ramesh03}
{Narayan}, R. \& {Heyl}, J.~S. 2003, \apj, 599, 419

\bibitem[{{{\"O}zel} {et~al.}(2016){{\"O}zel}, {Psaltis}, {G{\"u}ver}, {Baym},
  {Heinke}, \& {Guillot}}]{ozel16}
{{\"O}zel}, F., {Psaltis}, D., {G{\"u}ver}, T., {Baym}, G., {Heinke}, C., \&
  {Guillot}, S. 2016, \apj, 820, 28

\bibitem[{{Pandel} {et~al.}(2008){Pandel}, {Kaaret}, \& {Corbel}}]{pandel08}
{Pandel}, D., {Kaaret}, P., \& {Corbel}, S. 2008, \apj, 688, 1288

\bibitem[{{Paxton} {et~al.}(2015){Paxton}, {Marchant}, {Schwab}, {Bauer},
  {Bildsten}, {Cantiello}, {Dessart}, {Farmer}, {Hu}, {Langer}, {Townsend},
  {Townsley}, \& {Timmes}}]{mesa15}
{Paxton}, B., {Marchant}, P., {Schwab}, J., {Bauer}, E.~B., {Bildsten}, L.,
  {Cantiello}, M., {Dessart}, L., {Farmer}, R., {Hu}, H., {Langer}, N.,
  {Townsend}, R.~H.~D., {Townsley}, D.~M., \& {Timmes}, F.~X. 2015, \apjs, 220,
  15

\bibitem[{{Priedhorsky} \& {Terrell}(1984)}]{pt84}
{Priedhorsky}, W. \& {Terrell}, J. 1984, \apjl, 284, L17

\bibitem[{{Schatz} {et~al.}(2001){Schatz}, {Aprahamian}, {Barnard}, {Bildsten},
  {Cumming}, {Ouellette}, {Rauscher}, {Thielemann}, \& {Wiescher}}]{schatz01}
{Schatz}, H., {Aprahamian}, A., {Barnard}, V., {Bildsten}, L., {Cumming}, A.,
  {Ouellette}, M., {Rauscher}, T., {Thielemann}, F.-K., \& {Wiescher}, M. 2001,
  Physical Review Letters, 86, 3471

\bibitem[{{Schatz} {et~al.}(2003){Schatz}, {Bildsten}, \& {Cumming}}]{schatz03}
{Schatz}, H., {Bildsten}, L., \& {Cumming}, A. 2003, \apjl, 583, L87

\bibitem[{{Schatz} \& {Ong}(2016)}]{so16}
{Schatz}, H. \& {Ong}, W.-J. 2016, ArXiv e-prints 1610.07596

\bibitem[{{Steiner} {et~al.}(2013){Steiner}, {Lattimer}, \& {Brown}}]{slb13}
{Steiner}, A.~W., {Lattimer}, J.~M., \& {Brown}, E.~F. 2013, \apjl, 765, L5

\bibitem[{{Stella} {et~al.}(1987){Stella}, {White}, \& {Priedhorsky}}]{swp87}
{Stella}, L., {White}, N.~E., \& {Priedhorsky}, W. 1987, \apjl, 312, L17

\bibitem[{{Stevens} {et~al.}(2014){Stevens}, {Brown}, {Cumming}, {Cyburt}, \&
  {Schatz}}]{stevens14}
{Stevens}, J., {Brown}, E.~F., {Cumming}, A., {Cyburt}, R., \& {Schatz}, H.
  2014, \apj, 791, 106

\bibitem[{{Strohmayer} \& {Bildsten}(2006)}]{sb03}
{Strohmayer}, T. \& {Bildsten}, L. 2006, in Compact Stellar X-Ray Sources, ed.
  W.~H.~G. {Lewin} \& M.~{van der Klis} (Cambridge University Press),
  (astro--ph/0301544)

\bibitem[{{Strohmayer} \& {Brown}(2002)}]{stroh02}
{Strohmayer}, T.~E. \& {Brown}, E.~F. 2002, \apj, 566, 1045

\bibitem[{{Strohmayer} \& {Markwardt}(2002)}]{stroh02b}
{Strohmayer}, T.~E. \& {Markwardt}, C.~B. 2002, \apj, 577, 337

\bibitem[{{Tanaka}(1989)}]{tanaka89}
{Tanaka}, Y. 1989, in Proc. 23rd ESLAB Symposium on Two Topics in X-ray
  Astronomy, Bologna, Italy, 13-20 September, ed. J.~{Hunt} \& B.~{Battrick}
  No. SP-296 (ESTEC, Noordwijk, The Netherlands: ESA), 3--13

\bibitem[{{Thompson} {et~al.}(2008){Thompson}, {Galloway}, {Rothschild}, \&
  {Homer}}]{thompson08}
{Thompson}, T.~W.~J., {Galloway}, D.~K., {Rothschild}, R.~E., \& {Homer}, L.
  2008, \apj, 681, 506

\bibitem[{{Titarchuk}(1994)}]{tit94}
{Titarchuk}, L. 1994, \apj, 434, 570

\bibitem[{{Ubertini} {et~al.}(1999){Ubertini}, {Bazzano}, {Cocchi},
  {Natalucci}, {Heise}, {Muller}, \& {in 't Zand}}]{clock99}
{Ubertini}, P., {Bazzano}, A., {Cocchi}, M., {Natalucci}, L., {Heise}, J.,
  {Muller}, J.~M., \& {in 't Zand}, J.~J.~M. 1999, \apjl, 514, L27

\bibitem[{{van Paradijs} {et~al.}(1990){van Paradijs}, {van der Klis}, {van
  Amerongen}, {Pedersen}, {Smale}, {Mukai}, {Schoembs}, {Haefner}, {Pfeiffer},
  \& {Lewin}}]{1636orb}
{van Paradijs}, J., {van der Klis}, M., {van Amerongen}, S., {Pedersen}, H.,
  {Smale}, A.~P., {Mukai}, K., {Schoembs}, R., {Haefner}, R., {Pfeiffer}, M.,
  \& {Lewin}, W.~H.~G. 1990, \aap, 234, 181

\bibitem[{{Wijnands}(2001)}]{wij01b}
{Wijnands}, R. 2001, \apjl, 554, L59

\bibitem[{{Wijnands} \& {van der Klis}(1998)}]{wij98b}
{Wijnands}, R. \& {van der Klis}, M. 1998, \nat, 394, 344

\bibitem[{{Woosley} {et~al.}(2004){Woosley}, {Heger}, {Cumming}, {Hoffman},
  {Pruet}, {Rauscher}, {Fisker}, {Schatz}, {Brown}, \& {Wiescher}}]{woos04}
{Woosley}, S.~E., {Heger}, A., {Cumming}, A., {Hoffman}, R.~D., {Pruet}, J.,
  {Rauscher}, T., {Fisker}, J.~L., {Schatz}, H., {Brown}, B.~A., \& {Wiescher},
  M. 2004, \apjs, 151, 75

\bibitem[{{Worpel} {et~al.}(2013){Worpel}, {Galloway}, \& {Price}}]{worpel13a}
{Worpel}, H., {Galloway}, D.~K., \& {Price}, D.~J. 2013, \apj, 772, 94

\bibitem[{{Worpel} {et~al.}(2015){Worpel}, {Galloway}, \& {Price}}]{worpel15}
---. 2015, \apj, 801, 60

\bibitem[{{Zamfir} {et~al.}(2012){Zamfir}, {Cumming}, \&
  {Galloway}}]{zamfir12a}
{Zamfir}, M., {Cumming}, A., \& {Galloway}, D.~K. 2012, \apj, 749, 69

\bibitem[{{Zhang} {et~al.}(1997){Zhang}, {Lapidus}, {Swank}, {White}, \&
  {Titarchuk}}]{zhang97}
{Zhang}, W., {Lapidus}, I., {Swank}, J.~H., {White}, N.~E., \& {Titarchuk}, L.
  1997, \iaucirc, 6541

\end{thebibliography}
\end{document}